\documentclass[amsfonts,aps,nofootinbib,preprint]{revtex4}

\begin{document}

\title{An extension of the Linear Delta Expansion to Superspace}

\author{M. C. B. Abdalla{\footnote{mabdalla@ift.unesp.br}},
J. A. Helay\"{e}l-Neto{\footnote{helayel@cbpf.br}},
Daniel L. Nedel{\footnote{dnedel.unipampa@ufpel.tche.br}}
and Carlos R. Senise Jr.{\footnote{senise@ift.unesp.br}}}

\affiliation{Instituto de F\'{\i}sica Te\'{o}rica, UNESP, Rua Pamplona
145, Bela Vista, S\~{a}o Paulo, SP, 01405-900, Brazil }

\affiliation{Centro Brasileiro de Pesquisas F\'isicas, Rua Dr. Xavier Sigaud 150, 
Urca, Rio de Janeiro, RJ, 22290-180, Brazil}

\affiliation{Universidade Federal do Pampa, 
Rua Carlos Barbosa S/N, Bairro Get\'ulio Vargas, 96412-420, Bag\'e, RS, Brazil}

\begin{abstract}
We introduce and discuss the method of Linear Delta Expansion for the calculation of effective potentials in superspace, by adopting the improved version of the super-Feynman rules. Calculations are carried out up to two-loops and an expression for the optimized K\"{a}hler potential is worked out.
\end{abstract}

\maketitle
From among the most important tools of Quantum Field Theory (QFT) one selects the study of the effective action and the effective potential.
They have been used as a device to study important quantum properties of field theory, such as vacuum energy and symmetry breaking patterns.

There are well-known methods to compute effective actions and effective potentials to all finite orders in perturbation theory \cite{livro Shapiro}. Specially, the issue of effective potential calculations has been discussed in a wide variety of models, due to the relevance of the problem in proposing unified models for the fundamental interactions \cite{Salam-Strathdee,Lee,Weinberg}. Also, the important discussion of the gauge-(in)dependence of the effective potential in the framework of Yang-Mills theories has received a great deal of attention in connection with the calculation of physical quantities in the Electroweak Theory \cite{Ovrut 1,Ovrut 2,Fraser}. 

In the case of supersymmetric theories, a very powerful method to compute quantum corrections to the effective action is to use super-Feynman rules and then apply perturbation theory directly in superspace \cite{Miller,Helayel PL,Helayel NP,Srivastava}. In general, the super-effective action is described by two functions of the chiral and the antichiral superfields; one is required to be a holomorphic function and the other one, called K\"{a}hler potential, less constrained, is required to be just a real function. The superspace techniques allow us to use various non-renormalization theorems, which constrain the perturbative quantum corrections to the holomorphic part of the effective action and lead to all orders results \cite{Novikov,Intriligator1}. 
On the other hand, the K\"{a}hler potential encodes the wave function renormalization for the chiral superfields and receives corrections to all orders in perturbation theory. 

In general, the outstanding problems of QFT are typically non-perturbative. In this case it is necessary to develop a method to 
make possible a ressumation of Feynman diagrams and derive a result which has all orders of the coupling constant. Those methods sum infinite Feynman diagrams that belong to a specific set. The more traditional ressumation result is the Coleman-Weinberg potential \cite{CW}. It is the sum of all one-loop diagrams. There are many ways to derive the Coleman-Weinberg potential, the two traditional ones are the diagrammatic calculation and the functional calculation \cite{CW,Jac}. If the fields are defined in a N-dimensional representation of any group, the Coleman-Weinberg potential can also be derived from a 1/N expansion in the large N limit. If we are interested in finite N results, which means to make a ressumation that involves more than one-loop diagrams, both the diagrammatic and the functional methods are very difficult to use. Mainly because it is necessary to work with infinite diagrams, which turns the renormalization procedure a difficult task. 

Over the past years, an alternative ressumation method has been developed, namely the Linear Delta Expansion (LDE) \cite{delta1}. This method can easily reproduce the Coleman-Weinberg potential, and the use of the LDE in various QFT models has proved to be a powerful tool to derive finite N results \cite{Kneur}. The main characteristic of the method is to use a traditional perturbative approach together with an optimization procedure. So, in order to derive a non-perturbative result, it is just necessary to work with a few diagrams and use perturbative renormalization techniques.

 The main goal of this paper is to show that the LDE can be also a powerful method to derive non-perturbative results in supersymmetric theories. To this end, in Section I, we present the main steps of the method based on the LDE, in Section II, we further develop the method to be applied directly in superspace and, in Section III, we derive the Coleman-Weinberg potential plus two-loop corrections for the Wess-Zumino (WZ) model. Our Concluding Remarks are finally cast in Section IV.  

    \section{The Linear Delta Expansion}
    In this section we are going to make a brief review of the LDE. Starting with a Lagrangian ${\cal L}$, let us define the following ${\cal L}^{\delta}$ interpolated Lagrangian:
    \begin{equation}
    {\mathcal L}^{\delta}=\delta{\mathcal L}(\mu)+(1-\delta){\mathcal L}_{0}(\mu) \ , 
    \end{equation}
where $\delta$ is an arbitrary parameter, ${\cal L}_0(\mu)$ is the free Lagrangian and $\mu$ is a mass parameter. Note that when $\delta =1$  the original theory is retrieved, so $\delta$ is just a bookkeeping parameter. The $\delta$ parameter labels interactions and it is used as a perturbative coupling instead of the original coupling. The mass parameter appears in ${\cal L}_0$ and $\delta{\cal L}_0$. In fact, we are using the traditional trick consisting of adding and subtrac-\
ting a mass term in the original Lagrangian. The $\mu$ dependence of ${\cal L}_0$ is absorbed in the propagators, whereas $\delta{\cal L}_0$ is regarded as a quadratic interaction. 
 
 Let us now define the strategy of the method. We apply usual perturbation theory in $\delta$ and at the end we put $\delta =1$. 
Up to this stage 
traditional perturbation theory was applied, working with finite Feynman diagrams,
and the results are purely perturbative. However, quantities evaluated at finite order in $\delta$ depend explicitly on $\mu$. So, it is necessary to fix the $\mu$ parameter. There are two ways of doing it. The first one is to use the Principle of Minimal Sensitivity (PMS) \cite{PMS}. Since $\mu$ does not belong to the original theory, we may require that a physical quantity, such as the effective potential $V^{(k)}(\mu)$, calculated perturbatively to order $\delta ^k$, must be evaluated at a point where it is less sensitive to the parameter $\mu$. According to the PMS  $\mu={\mu_0}$ is the solution of the equation
 \begin{equation}
 \left.\frac{\partial V^{(k)}(\mu)}{\partial\mu}\right|_{\mu=\mu_0,\delta=1}=0  . \label{PMS}
 \end{equation}
After this procedure, the optimum value ${\mu_0}$ will be a function of the original coupling and the fields. Then we replace ${\mu_0}$ into the effective potential $V^{(k)}$ and obtain a non-perturbative result, since the propagator depends on $\mu$. 
 
 The second way to fix $\mu$ is known as the Fastest Apparent Convergence (FAC) criterion \cite{PMS}. It requires that from any k coefficient of the perturbative expansion
 
 \begin{equation}
 V^{(k)}(\mu)=\sum_{i=0}^{k}c_i(\mu)\delta^i \label{coef} \ ,
 \end{equation}
 
\noindent the following relation is fulfilled:   
 
 \begin{equation}
 \left.\left[V^k(\mu)- V^{k-1}(\mu)\right]\right|_{\delta=1}=0 . \label{fac}
 \end{equation}

Again, the ${\mu_0}$ solution of the above equation will be a function of the original couplings and the fields, and when we replace $\mu={\mu_0}$ into $V(\mu)$ we obtain a non-perturbative result. The equation (\ref{fac}) is equivalent to taking the k-th coefficient of equation (\ref{coef}) equals to zero ($c_k=0$). If we are interested in an order $\delta^{k}$ result ( $V^{(k)}(\mu)$) using the FAC criterion, it is just necessary to find the solution of the equation $\displaystyle\left.c_{k+1}(\mu)\right|_{\mu=\mu_0}=0$ and put it in $V^{(k)}(\mu)$. The reference \cite{Kneur} provides an extensive list of successful applications of the method. 

To exemplify the above results, we calculate the effective potential for the Gross-Neveu model \cite{G-N} 
at order $\delta$. This is a model consisting in $N$ fermions with quartic interaction, in $(1+1)$-dimension, and it is interesting for the study of some important subjects in field theory, such as discrete chiral symmetry breaking, dimensional transmutation and assymtoptic freedom. It was first analyzed with LDE in \cite{MG} ,where the large N result was reproduced. Important finite N results was derived  in \cite{MRE1,MRE2}, where the emergence of a tricritical point was discovered.   

    The original Lagrangian of the model is
\begin{equation}
{\mathcal L}=i\bar{\psi}^{a}\partial_{m}\gamma^{m}\psi^{a}+\frac{g}{2N}(\bar{\psi}^{a}\psi^{a})^{2}\ , \label{1/N escalar 1}
\end{equation}
with $a=1,2,...,N$. 
We now introduce the auxiliary field $\sigma$ in the form
\begin{equation}
{\mathcal L}\!\rightarrow\!{\mathcal L}-\frac{N}{2g}\left(\sigma-\frac{g}{N}\bar{\psi}^{a}\psi^{a}\right)^{2} \ , \label{1/N escalar 2}
\end{equation}
which allows us to make a $1/N$ expansion. Solving the Euler-Lagrange equations, we have
\begin{equation}
\sigma=\frac{g}{N}\bar{\psi}^{a}\psi^{a} \ , \label{1/N escalar 3}
\end{equation}
which is a constraint equation. Thus, the new Lagrangian is
\begin{eqnarray}
{\mathcal L}&\rightarrow&{\mathcal L}+\mu\bar{\psi}^{a}\psi^{a}-\mu\bar{\psi}^{a}\psi^{a} \nonumber\\
&=&i\bar{\psi}^{a}\partial_{m}\gamma^{m}\psi^{a}-\frac{N}{2g}\sigma^{2}+\sigma\bar{\psi}^{a}\psi^{a}+\mu\bar{\psi}^{a}\psi^{a}-\mu\bar{\psi}^{a}\psi^{a} \nonumber\\
&=&{\mathcal L}_{0}(\mu)+{\mathcal L}_{int}(\mu) \ ,
\end{eqnarray}
with
\begin{eqnarray}
{\mathcal L}_{0}(\mu)&=&i\bar{\psi}^{a}\partial_{m}\gamma^{m}\psi^{a}-\frac{N}{2g}\sigma^{2}-\mu\bar{\psi}^{a}\psi^{a} \ , \nonumber\\
{\mathcal L}_{int}(\mu)&=&\sigma\bar{\psi}^{a}\psi^{a}+\mu\bar{\psi}^{a}\psi^{a} \ .
\end{eqnarray}
The interpolated Lagrangian is
\begin{eqnarray}
{\mathcal L}^{\delta}&=&\delta{\mathcal L}(\mu)+(1-\delta){\mathcal L}_{0}(\mu)  \nonumber\\
&=&i\bar{\psi}^{a}\partial_{m}\gamma^{m}\psi^{a}-\frac{N}{2g}\sigma^{2}-\mu\bar{\psi}^{a}\psi^{a}+\delta\sigma\bar{\psi}^{a}\psi^{a}+\delta\mu\bar{\psi}^{a}\psi^{a} \ . 
\end{eqnarray}
In this expression, the $-\mu\bar{\psi}^{a}\psi^{a}$ term is a mass term appearing in the propagator and the term $\delta\mu\bar{\psi}^{a}\psi^{a}$ represents an interaction with weight $\delta$. 
Note that the $\sigma$-propagator carries the $1/N$-factor and each fermion-loop contains a $N$ factor.

In general,  when the effective potential is calculated in 
QFT we do not worry about vacuum diagrams, since they do not depend on fields. However, the vacuum diagrams depend on $\mu$ and are important in the LDE, since the arbitrary mass parameter will depend on fields after the optimization procedure. So, in the LDE, it is necessary to calculate the vacuum diagrams order by order. 

\newpage
The diagrammatic representation of the effective potential at order $\delta $ is
\begin{eqnarray}
 \begin{picture}(280,5)
\put(41,-0,5){\line(50,0){4}}\put(47,-0,5){\line(50,0){4}}\put(53,-0,5){\line(50,0){4}}\put(59,-0,5){\line(50,0){4}}\put(65,-0,5){\line(50,0){4}}\put(71,-0,5){\line(50,0){4}}\put(77,-0,5){\line(50,0){4}}\put(83,-0,5){\line(50,0){4}}
\put(92,-3){+}
\put(118,0){\circle{25}}\put(135,-3){+}
\put(150,-0,5){\line(50,0){4}}\put(156,-0,5){\line(50,0){4}}\put(162,-0,5){\line(50,0){4}}\put(165.7,-3){$\bullet$}\put(180,0){\circle{25}}\put(197,-3){+}
\put(223,0){\circle{25}}\put(230,2){$\bullet$} 
 \end{picture},
 \label{order1}
 \end{eqnarray}
 \hspace{1,7cm} Figure 1: Effective potential for the Gross-Neveu model at order $\delta$.

\vspace{1cm}
\noindent where the dashed line represents the $\sigma$-propagator. The second and the last diagrams are the vacuum contribution at order zero and at order $\delta$. Using the Feynman rules, the diagrammatic sum corresponds to :
\begin{eqnarray}
V_{eff}(\sigma_{c})= \frac{N}{2g}\sigma_{c}^{2}+iN\int\!\frac{d^{2}p}{(2\pi)^{2}}\ln(p^{2}-\mu^{2})-2iN\delta \int\!\frac{d^{2}p}{(2\pi)^{2}}\frac{\mu\sigma_{c}}{(p^{2}-\mu^{2})}+ \nonumber\\
+2iN\delta\int\!\frac{d^{2}p}{(2\pi)^{2}}\frac{\mu^{2}}{(p^{2}-\mu^{2})} \ , \hspace{6cm} \label{Veff sigma}
\end{eqnarray}
where $\sigma_{c}$ is the classical field. The above expression corresponds to the classical potential ($N\sigma_{c}^{2}/2g$) plus the one-loop correction for the vacuum and  one-point Green's function at order $\delta$. 

Now we apply the PMS at order $\delta$ and take $\delta=1$. We impose
\begin{equation}
\left.\frac{\partial V_{eff}}{\partial\mu}\right|_{\mu=\mu_{0}}=0 \ ,
\end{equation}
which  implies
\begin{equation}
\mu_{0}=\sigma_{c} \ .
\end{equation}
Substituting this result in (\ref{Veff sigma}), the effective potential evaluated at $\mu=\mu_{0}$ is
\begin{equation}
\frac{V_{eff}(\sigma_{c})}{N}=\frac{1}{2g}\sigma_{c}^{2}+i\int\!\frac{d^{2}p}{(2\pi)^{2}}\ln\left(1-\frac{\sigma_{c}^{2}}{p^{2}+i\varepsilon}\right) \ , \label{Veff mu til}
\end{equation}
which corresponds to the effective potential of the Gross-Neveu model in the $N\!\rightarrow\!\infty$ limit \cite{SidCol 2}, and represents the sum of all one-loop diagrams for this theory. This result shows the simplicity of the method. We can reproduce an infinite sum of a class of diagrams, working with few diagrams, represented in (\ref{order1}).

\section{LDE in superspace}
 
 Let us now further develop the LDE for superspace applications. We are going to use the WZ model. The superspace action is
 \begin{equation}
 S[\Phi, \bar\Phi]=\int d^{8}z\bar\Phi \Phi + \int d^{6}z \left(\frac{m}{2}\Phi^{2} + \frac{\lambda}{3!}\Phi^{3}\right) + \int d^{6}\bar{z}\left(\frac{m}{2}\bar{\Phi}^{2} + \frac{\bar{\lambda}}{3!}\bar{\Phi}^{3}\right) ,
 \end{equation}
where we have the notation $d^{8}z=d^{4}x d^{2}\theta d^{2}\bar\theta$, $d^{6}z=d^{4}x d^{2}\theta$, $d^{6}\bar{z}=d^{4}x d^{2}\bar\theta$. Using the non-renormalization theorems, the more general effective action for the WZ model is written as
\begin{eqnarray}
\Gamma[\Phi,\bar{\Phi}]=\int d^{8}z\left[K(\bar{\Phi},\Phi)+... 
\right] \ \label{pw1}  
\end{eqnarray}  

\noindent where the first term is the K\"{a}hler Potential and the 
other terms involve derivatives of the fields.
The effective action can be calculated perturbatively directly in superspace using superspace Feynman rules. Here we use the improved Feynman rules defined originally in \cite{Grisaru}. 
 
 Let us start to apply the LDE in the WZ model. The first modification is to implement two mass parameters, $\mu$ and $\bar{\mu}$, instead of just one. In order to fix these parameters we use two optimization equations. In particular we use the FAC criterion, so that we have one equation similar to (\ref{fac}) for $\mu$ and other for $\bar{\mu}$. Using two mass parameters the Lagrangian ${\cal L}^\delta$  can be written as
\begin{eqnarray}
{\mathcal L}^{\delta}=\delta{\mathcal L}+(1-\delta){\mathcal L}_{0} \hspace{5,65cm} \nonumber\\
=\int\!d^{4}\theta\bar{\Phi}\Phi+\int\!d^{2}\theta\left(\frac{M}{2}\Phi^{2}+\frac{\delta\lambda}{3!}\Phi^{3}-\frac{\delta\mu}{2}\Phi^{2}\right)+ \nonumber\\
+\int\!d^{2}\bar{\theta}\left(\frac{\bar{M}}{2}\bar{\Phi}^{2}+\frac{\delta\bar{\lambda}}{3!}\bar{\Phi}^{3}-\frac{\delta\bar{\mu}}{2}\bar{\Phi}^{2}\right),  \label{linter}
\end{eqnarray} 
 where $M= m +\mu$ and $\bar{M} = m+\bar{\mu}$. The action ${\cal S}^\delta$ can be written using the matrix notation as \cite{Buckl}
 \begin{eqnarray}
 S^{\delta}[\Phi,\bar{\Phi}]&=&\frac{1}{2}\int\!dz \ dz^{\prime}\left(\Phi(z) \ \ \bar{\Phi}(z)\right)
H^{(M,\bar{M})}
\left( \begin{array}{c}
\Phi(z^{\prime}) \\
\bar{\Phi}(z^{\prime})
\end{array}
\right) \nonumber \\ &+&\int\!d^{6}z\left(\frac{\delta\lambda}{3!}\Phi^{3}-\frac{\delta\mu}{2}\Phi^{2}\right)+\int\!d^{6}\bar{z}\left(\frac{\delta\bar{\lambda}}{3!}\bar{\Phi}^{3}-\frac{\delta\bar{\mu}}{2}\bar{\Phi}^{2}\right), \label{acdelta}
\end{eqnarray}
where
 \begin{equation}
 H^{(M,\bar{M})}=\left( \begin{array}{cc}
M & -\frac{1}{4}\bar{D}^{2} \\
-\frac{1}{4}D^{2} & \bar{M}
\end{array}
\right)\left( \begin{array}{cc}
\delta_{+}(z,z^{\prime}) & 0 \\
0 & \delta_{-}(z,z^{\prime})
\end{array}
\right)  . \label{H} 
\end{equation}
Now one has a new chiral and antichiral quadratic interaction proportional to $\delta\mu$ and $\delta\bar{\mu}$. Also, the superpropagator $G^{M,\bar{M}}$, defined in terms of the chiral and antichiral delta functions ($\delta_{+}(z,z^{\prime}),\delta_{-}(z,z^{\prime}))$ by the equation
 \begin{equation}
  G^{(M,\bar{M})}H^{(M,\bar{M})}= \left( \begin{array}{cc}
\delta_{+}(z,z^{\prime}) & 0 \\
0 & \delta_{-}(z,z^{\prime})
\end{array}
\right) \ ,
  \end{equation}
has shifted mass terms $M= m +\mu$ and $\bar{M} = m+\bar{\mu}$. The new Feynman rules can now be written as:
  
\begin{itemize}
  \item 1. Propagators:
  \begin{eqnarray}
  \Phi\bar{\Phi}&:&-\frac{i}{k^{2}+M\bar{M}}\delta^{4}(\theta-\theta^{\prime}) \ , \nonumber\\
  \Phi\Phi&:&\frac{iM}{k^{2}(k^{2}+M\bar{M})}\frac{1}{4}D^{2}(k)\delta^{4}(\theta-\theta^{\prime}) \ , \\ 
\bar{\Phi}\bar{\Phi}&:&\frac{i\bar{M}}{k^{2}(k^{2}+M\bar{M})}\frac{1}{4}\bar{D}^{2}(k)\delta^{4}(\theta-\theta^{\prime}) \ ; \nonumber
  \end{eqnarray}
  
  \item 2. Vertices:
  \begin{eqnarray}
  \bullet=i\delta\lambda\int d^{8}z \ , \ \ \ \ \ \ \ \circ=i\delta\bar{\lambda}\int d^{8}z \ , \nonumber
  \end{eqnarray}
\begin{eqnarray}
\begin{picture}(101,5)
\thicklines \put(50,2){\line(50,0){50}}\put(72,-0,3){$\times$}
\end{picture}
=-i\delta\int\mu d^{8}z \ , \hspace{1,7cm}
\end{eqnarray}
\begin{eqnarray}
\begin{picture}(101,5)
\thicklines \put(50,2){\line(50,0){50}}\put(72,-0,3){$\otimes$}
\end{picture}
=-i\delta\int\bar{\mu}d^{8}z \ . \nonumber \hspace{1,7cm}
\end{eqnarray}
\end{itemize}

The perturbative effective K\"{a}hler potential can be calculated in powers of $\delta$ using the OPI functions. 
It is well-known that, in superspace, vacuum superdiagrams are identically zero, owing to Berezin integrals. To avoid this, we have to consider, from the very beginning, the parameters $\mu$, $\bar{\mu}$ as superfields and keep the vacuum supergraphs until the optimization procedure. From the supergenerator functional in the presence of the chiral ($J$) and antichiral ($\bar J$) sources:
 \begin{equation}
 \tilde{Z}[J,\bar{J}]=exp\left[iS_{INT}\left(\frac{1}{i}\frac{\delta}{\delta J},\frac{1}{i}\frac{\delta}{\delta\bar{J}}\right)\right]exp\left[\frac{i}{2}
(J,\bar{J})G^{(M,\bar{M})}\left( \begin{array}{c}
J \\
\bar{J}
\end{array}
\right)\right] \ ,
 \end{equation}
\noindent we can write the super-effective action:
\begin{equation}
\Gamma[\Phi,\bar{\Phi}]=-\frac{i}{2}\ln[sDet\left(G^{(M,\bar{M})}\right)]-i\ln\tilde{Z}[J,\bar{J}]-\int\!d^{6}zJ(z)\Phi(z)-\int\!d^{6}\bar{z}\bar{J}(z)\bar{\Phi}(z),             \label{sea}
 \end{equation}
where $sDet\left(G^{(M,\bar{M})}\right)$ is the superdeterminant of the matrix propagator, which in general is equal to one, but here we keep it because  $ G^{(M,\bar{M})}$ depends on $\mu$ and $\bar{\mu}$. Also, due to the $\mu$ and $\bar{\mu}$ dependence, the vacuum diagrams supergenerator $\tilde{Z}[0,0]$ is  not identically equal to one. We can define the normalized functional generator as $Z_N = \frac{\tilde{Z}[J,\bar J]}{\tilde{Z}[0,0]}$ and write the effective action as
 \begin{equation}
 \Gamma[\Phi,\bar{\Phi}]=-\frac{i}{2}\ln[sDet(G)]-i\ln\tilde{Z}[J_{0},\bar{J}_{0}]+\Gamma_{N}[\Phi,\bar{\Phi}] \ , \label{Gamma Phi barPhi}
 \end{equation}
where the sources $J_0$ and $\bar{J}_0$ are defined by the equations 
\begin{eqnarray}
\frac{\delta W[J,\bar{J}]}{\delta J(z)}|_{J=J_{0}}=\frac{\delta W[J,\bar{J}]}{\delta \bar{J}(z)}|_{\bar{J}=\bar{J}_{0}}=\frac{\delta\tilde{Z}[J,\bar{J}]}{\delta J(z)}|_{J=J_{0}}=\frac{\delta\tilde{Z}[J,\bar{J}]}{\delta\bar{J}(z)}|_{\bar{J}=\bar{J}_{0}}=0 \ . \label{cap5 6}
\end{eqnarray}
In equation (\ref{Gamma Phi barPhi}) the first two terms represent the vacuum diagrams (which are usually zero) and $\Gamma_N[\Phi,\bar{\Phi}]$ is the usual contribution to the effective action.

 \section{ K\"{a}hler Potential in the LDE using the FAC criterion}
 Let us now calculate the K\"{a}hler potential using the LDE up to the order $\delta^{2}$. We are going to show that this corresponds to the Coleman-Weinberg potential plus a sum of infinite two-loop supergraphs. 

 In figure 2 one  can see the diagrammatic sum of the effective K\"{a}hler potential up to the order $\delta^{2}$ (${\mathcal V}_{eff}^{\delta^{2}}$).
The tad-pole diagrams are zero, as usual, since they have quadratic terms of Grassmann delta functions.
  
 \begin{eqnarray}
 \begin{picture}(400,5)\thicklines 
\put(50,0){\circle{25}}\put(70,-3){+}
\put(85,-0,5){\line(50,0){18}}\put(100.7,-3){$\bullet$}\put(115,0){\circle{25}}\put(123,-3.2){$\otimes$}\put(135,-3){+}
\put(150,-0,5){\line(50,0){18}}\put(165.7,-3){$\circ$}\put(180,0){\circle{25}}\put(188,-3.2){$\times$}\put(200,-3){+}
\put(225,0){\circle{25}}\put(221,10){$\otimes$}\put(221.5,-14.5){$\times$}\put(242,-3){+}
\put(255,-0.5){\line(50,0){18}}\put(271,-3){$\circ$}\put(285,0){\circle{25}}\put(294,-3){$\bullet$}\put(297,-0.5){\line(50,0){18}}\put(320,-3){+}
\put(344,0){\circle{25}}\put(344,-12){\line(0,5){24}}\put(341.4,9.5){$\circ$}\put(341.5,-14){$\bullet$}
\end{picture}
 \label{order2}
 \end{eqnarray}
 \hspace{2,65cm} Figure 2: One and two-loop diagrams up to the order $\delta^{2}$. 

\vspace{1cm}
The diagrammatic sum of figure 2 corresponds to the terms: 
\begin{eqnarray}
{\mathcal V}_{eff}^{\delta^{2}}=&-&\frac{i}{2}sTr\ln\left[H^{(M,\bar{M})}\right]-\frac{i}{2}\delta^{2}\lambda\int\frac{d^{4}k}{(2\pi)^{4}}\int\!d^{4}\theta\frac{\bar{\mu}\Phi(0,\theta)}{(k^{2}+M\bar{M})^{2}}+ \nonumber\\
&-&\frac{i}{2}\delta^{2}\bar{\lambda}\int\frac{d^{4}k}{(2\pi)^{4}}\int\!d^{4}\theta\frac{\mu\bar{\Phi}(0,\theta)}{(k^{2}+M\bar{M})^{2}}+\frac{i}{2}\delta^{2}\int\frac{d^{4}k}{(2\pi)^{4}}\int\!d^{4}\theta\frac{\mu\bar{\mu}}{(k^{2}+M\bar{M})^{2}}+ \nonumber\\
&+&\frac{i}{2}\delta^{2}\lambda\bar{\lambda}\!\int\!\frac{d^{4}k}{(2\pi)^{4}}\!\int\!d^{4}\theta\frac{\bar{\Phi}(0,\theta)\Phi(0,\theta)}{(k^{2}+M\bar{M})^{2}}\!+\!\frac{1}{6}\delta^{2}\lambda\bar{\lambda}\!\int\!\frac{d^{4}kd^{4}q}{(2\pi)^{8}}\!\int\!d^{4}\theta\frac{1}{A} \ , \label{ordem delta 2} 
\end{eqnarray}
where $A=(k^{2}+M\bar{M})(q^{2}+M\bar{M})((q-k)^{2}+M\bar{M})$. We wrote the first term of equation (\ref{sea}) as a supertrace in the full superspace (see ref \cite{Buckl} for details). Now we have to fix the mass parameters and write the K\"{a}hler potential of equation (\ref{order2}) at $\delta =1$ in terms of the optimized parameters $\mu_0$ and $\bar{\mu}_0$. The FAC criterion is employed as an optimization procedure. In order to calculate the K\"{a}hler effective potential up to order 2 in $\delta$ (${\mathcal V}_{eff}^{\delta^{2}}$) we have to solve for $\mu_0$ and $\bar{\mu}_0$ the equation
 \begin{eqnarray}
 c^3(\mu,\bar\mu)&=&0 \ , \label{order3}
\end{eqnarray}
 at $\delta =1$, where $c^3(\mu,\bar\mu)$ corresponds to the $\delta^{3}$ coefficients in the perturbative expansion of the K\"{a}hler potential. 

The order $\delta^3$ (${\mathcal V}_{eff}^{\delta^{3}}$) diagrams are drawn in figure 3 below:

\begin{eqnarray*}
\begin{picture}(350,5) \thicklines
\put(15,0){\line(50,0){18}}\put(45,0){\circle{25}}\put(31,-2.2){$\circ$}\put(51,-10.5){$\bullet$}\put(50.5,6){$\bullet$}\put(64,-18.3){\line(-1,1){10}}\put(53.3,8.8){\line(1,1){10}}\put(65,-3){+}
\put(77,-0.5){\line(50,0){15}}\put(90,-2.5){$\circ$}\put(104,0){\circle{25}}\put(100,9.5){$\times$}\put(113.5,-2.5){$\bullet$}\put(116,-0.5){\line(50,0){15}}\put(134,-3){+}
\put(145,0){\line(50,0){18}}\put(175,0){\circle{25}}\put(161,-2.2){$\circ$}\put(171.2,-14){$\times$}\put(171,10){$\times$}\put(190,-3){+}
\put(214,0){\circle{25}}\put(198,-2.5){$\otimes$}\put(210.2,9.8){$\times$}\put(210.2,-14.5){$\times$}\put(228,-3){+}
\put(238,0){\line(50,0){18}}\put(268,0){\circle{25}}\put(253.6,-2.7){$\bullet$}\put(264,9.8){$\otimes$}\put(264,-14.2){$\times$}\put(282,-3){+}
\put(294,0){\line(50,0){18}}\put(324,0){\circle{25}}\put(336,0){\line(50,0){18}}\put(309.6,-2.7){$\bullet$}\put(333.3,-2.7){$\bullet$}\put(320,9.8){$\otimes$} 
\end{picture}
\end{eqnarray*}

\begin{eqnarray*}
\begin{picture}(350,5) \thicklines
\put(15,0){\line(50,0){18}}\put(45,0){\circle{25}}\put(31,-2.2){$\bullet$}\put(51,-10.5){$\circ$}\put(50.5,6){$\circ$}\put(64,-18.3){\line(-1,1){10}}\put(53.3,8.8){\line(1,1){10}}\put(65,-3){+}
\put(77,-0.5){\line(50,0){15}}\put(90,-2.5){$\bullet$}\put(104,0){\circle{25}}\put(100,9.5){$\otimes$}\put(113.5,-2.5){$\circ$}\put(116,-0.5){\line(50,0){15}}\put(134,-3){+}
\put(145,0){\line(50,0){18}}\put(175,0){\circle{25}}\put(161,-2.2){$\bullet$}\put(171.2,-14){$\otimes$}\put(171,10){$\otimes$}\put(190,-3){+}
\put(214,0){\circle{25}}\put(198,-2.5){$\times$}\put(210.2,9.8){$\otimes$}\put(210.2,-14.5){$\otimes$}\put(228,-3){+}
\put(238,0){\line(50,0){18}}\put(268,0){\circle{25}}\put(253.6,-2.7){$\circ$}\put(264,9.8){$\times$}\put(264,-14.2){$\otimes$}\put(282,-3){+}
\put(294,0){\line(50,0){18}}\put(324,0){\circle{25}}\put(336,0){\line(50,0){18}}\put(309.6,-2.7){$\circ$}\put(333.3,-2.7){$\circ$}\put(320,9.8){$\times$} 
\end{picture}
\end{eqnarray*}

\begin{eqnarray*}
\begin{picture}(350,5) \thicklines
\put(15,0){\line(50,0){18}}\put(45,0){\circle{25}}\put(30.8,-2.5){$\bullet$}\put(45,-12){\line(0,5){24}}\put(42.7,-14.5){$\circ$}\put(42.7,9.5){$\bullet$}\put(60,-3){+}
\put(70,0){\line(50,0){18}}\put(100,0){\circle{25}}\put(85.7,-2.5){$\circ$}\put(100,-12){\line(0,5){24}}\put(97.7,-14.5){$\circ$}\put(97.7,9.5){$\bullet$}\put(115,-3){+}
\put(139,0){\circle{25}}\put(123,-2.5){$\times$}\put(139,-12){\line(0,5){24}}\put(136.5,-14.5){$\circ$}\put(136.5,9.5){$\bullet$}\put(154,-3){+}
\put(179,0){\circle{25}}\put(163.2,-2.5){$\otimes$}\put(179,-12){\line(0,5){24}}\put(176.5,-14.5){$\circ$}\put(176.5,9.5){$\bullet$}\put(194,-3){+}
\put(217,0){\circle{25}}\put(213.2,-2.5){$\times$}\put(217,-12){\line(0,5){24}}\put(214.4,-14.5){$\circ$}\put(214.4,9.5){$\bullet$}\put(231.9,-3){+}
\put(255,0){\circle{25}}\put(251.4,-2.5){$\otimes$}\put(255,-12){\line(0,5){24}}\put(252.5,-14.5){$\circ$}\put(252.5,9.5){$\bullet$}\put(269.8,-3){+}
\put(293,0){\circle{25}}\put(290.5,-2.5){$\bullet$}\put(293,-12){\line(0,5){24}}\put(290.6,-14.5){$\circ$}\put(290.6,9.5){$\bullet$}\put(292.5,-0.4){\line(1,0){10}}\put(308,-3){+}
\put(331,0){\circle{25}}\put(328.6,-2.5){$\circ$}\put(331,-12){\line(0,5){24}}\put(328.6,-14.5){$\circ$}\put(328.5,9.5){$\bullet$}\put(331,-0.2){\line(1,0){10}}
\end{picture}
\end{eqnarray*}
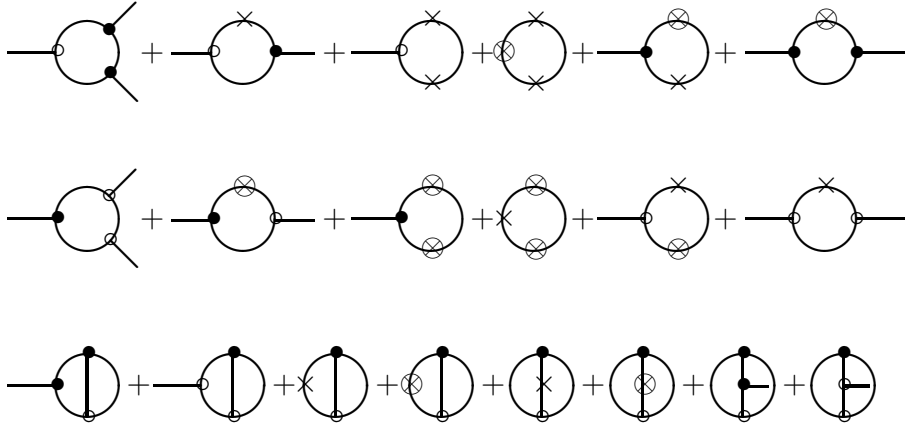
\hspace{3cm} Figure 3: One and two-loop diagrams up to the order $\delta^{3}$.

\vspace{1cm} 
Using the Feynman rules, they correspond to the following term of the K\"{a}hler potential: 
\begin{eqnarray}
\frac{i}{2}\delta^{3}\!\int\!\frac{d^{4}k}{(2\pi)^{4}}\!\int \!d^{4}\theta\frac{1}{(k^{2}\!+\!M\bar{M})^{3}}\!\left[\!-\lambda^{2}\bar{\lambda}M\Phi^{2}\bar{\Phi}\!+\!\lambda\bar{\lambda}M\mu\Phi\bar{\Phi}\!-\!\bar{\lambda}M\mu^{2}\bar{\Phi}\!+\!M\mu^{2}\bar{\mu}\!-\!\lambda M\mu\bar{\mu}\Phi+\!\right. \hspace{0,25cm} \nonumber\\
\left.+\lambda^{2}M\bar{\mu}\Phi^{2}\!-\!\lambda\bar{\lambda}^{2}\bar{M}\Phi\bar{\Phi}^{2}\!+\!\lambda\bar{\lambda}\bar{M}\bar{\mu}\Phi\bar{\Phi}\!-\!\lambda\bar{M}\bar{\mu}^{2}\Phi\!+\!\bar{M}\mu\bar{\mu}^{2}\!-\!\bar{\lambda} \bar{M}\mu\bar{\mu}\bar{\Phi}\!+\!\bar{\lambda}^{2}\bar{M}\mu\bar{\Phi}^{2}\right]\!+ \hspace{0,3cm} \nonumber\\
+\frac{1}{4}\delta^{3}\lambda\bar{\lambda}\!\!\int\!\!\frac{d^{4}kd^{4}q}{(2\pi)^{8}}\!\!\int \!\!d^{4}\theta\!\!\left[\!\frac{1}{B}\!\left(\!-\lambda M\Phi\!-\!\bar{\lambda}\bar{M}\bar{\Phi}\!+\!\mu M\!+\!\bar{\mu}\bar{M}\right)\!+\!\frac{1}{C}\!\left(\mu M\!+\!\bar{\mu}\bar{M}\!-\!\lambda M\Phi\!-\!\bar{\lambda}\bar{M}\bar{\Phi}\right)\!\right] \ , \hspace{0,2cm}
\end{eqnarray}
where each term corresponds to its respective diagram in figure 3, in the same order. Rearranging the terms in the above equation we obtain that ${\mathcal V}_{eff}^{\delta^{3}}$ corresponds to
\begin{eqnarray}
\frac{i}{2}\delta^{3}\!\int\!\frac{d^{4}k}{(2\pi)^{4}}\!\int \!d^{4}\theta\frac{1}{(k^{2}\!+\!M\bar{M})^{3}}\!\left[\lambda\bar{\lambda}M\Phi\bar{\Phi}(\!-\lambda\Phi\!+\!\mu)\!+\!M\mu^{2}(\!-\bar{\lambda}\bar{\Phi}\!+\!\bar{\mu})\!+\!\lambda M\bar{\mu}\Phi(\!-\mu\!+\!\lambda\Phi)+\right. \hspace{0cm} \nonumber\\
\left.+\lambda\bar{\lambda}\bar{M}\Phi\bar{\Phi}(\!-\bar{\lambda}\bar{\Phi}\!+\!\bar{\mu})\!+\!\bar{M}\bar{\mu}^{2}(\!-\lambda\Phi\!+\!\mu)\!+\!\bar{\lambda}\bar{M}\mu\bar{\Phi}(\!-\bar{\mu}\!+\!\bar{\lambda}\bar{\Phi})\right]\!+ \hspace{0cm} \nonumber\\
+\frac{1}{4}\delta^{3}\lambda\bar{\lambda}\!\!\int\!\!\frac{d^{4}kd^{4}q}{(2\pi)^{8}}\!\!\int \!\!d^{4}\theta\!\!\left[\!\frac{1}{B}\!\left(\!M(\!-\lambda\Phi\!+\!\mu)\!+\!\bar{M}(\!-\bar{\lambda}\bar{\Phi}\!+\!\bar{\mu})\right)\!+\!\frac{1}{C}\left(\!M(\mu\!-\!\lambda\Phi)\!+\!\bar{M}(\bar{\mu}\!-\!\bar{\lambda}\bar{\Phi})\right)\right] \ , \hspace{0,4cm} \label{ordem delta 3} 
\end{eqnarray}
where $C =(k^{2}+M\bar{M})(q^{2}+M\bar{M})((q-k)^{2}+M\bar{M})^2 $ and $B=(k^{2}+M\bar{M})^2(q^{2}+M\bar{M})((q-k)^{2}+M\bar{M})$. From the above equation we can derive the following simple solution for equation (\ref{order3}), before calculating the integrals:
\begin{eqnarray}
\mu_0 &=& \lambda\Phi \ , \nonumber \\
\bar{\mu}_0&=& \bar{\lambda}\bar{\Phi} \ .
\end{eqnarray}

Finally, substituting this solution into equation (\ref{ordem delta 2}) at $\delta =1$ we have only two terms for the optimized K\"{a}hler potential:
\begin{eqnarray}
{\mathcal V}_{eff}=&-&\frac{i}{2}sTr\ln\left[H^{(M_0,\bar{M}_0)}\right] + \nonumber \\ &+&\frac{1}{6}\lambda\bar{\lambda}\!\int\!\frac{d^{4}kd^{4}q}{(2\pi)^{8}}\!\int\!d^{4}\theta\frac{1}{(k^{2}+M_0\bar{M}_0)(q^{2}+M_0\bar{M}_0)((q-k)^{2}+M_0\bar{M}_0)} \ ,
\end{eqnarray}
where $H^{(M_0,\bar{M}_0)}$ is the same as $H^{(M,\bar{M})}$ defined in (\ref{H}) replacing $M$ by $M_0= m+ \lambda\Phi$ and $\bar{M}$ by 
$\bar{M}_0 = m+ \bar{\lambda}\bar{\Phi}$. The first one corresponds to the Coleman-Weinberg potential and it represents the sum of all one-loop supergraphs. The second corresponds to a sum of infinite Feynman diagrams that belong to a special set of two-loop diagrams. We can now regularize the integrals and go ahead by adopting renormalization approaches compatible with supersymmetry \cite{livro Grisaru}.

\section{Concluding Remarks}
We have applied superspace and supergraph techniques to carry out the LDE for the WZ model. We understand that this is a first step towards a more substantial programme of applications of the efficacy of supergraph methods to compute quantum corrections in the framework of supersymmetric field theories.

The next natural systems to probe the method are the O'Raifeartaigh \cite{O'Raifeartaigh} and Fayet-Iliopoulos \cite{Fayet-Iliopoulos} models, which realise the spontaneous breaking of supersymmetry. We have interesting non-renormalisation theorems \cite{livro Grisaru} to be satisfied and these calculations may be a good test for the LDE in the situation of spontaneous breaking of supersymmetry. 

More relevant for phenomenology are the models in which the soft explicitly breaking terms are introduced. The latter have been carefully classified and studied by Girardello and Grisaru, in the work of reference \cite{Girardello-Grisaru}. With those (phenomenologically interesting) terms, supergraph techniques become less obvious, but they have anyhow been extended to include the effect of the supersymmetry breaking parameters to all orders \cite{Helayel PL}. We believe that the application of the LDE to such a class of models may be relevant for the understanding of the method and also for the sake of application in phenomenologically realistic models based on supersymmetry. We shall be reporting on these results in a forthcoming work.

Finally, we point out that it would be of interest to investigate the control of the convergence of the LDE in superspace. This question has not yet been adressed to, but its development could bring a new insight and new elements in the extension of the LDE to superspace.

 \end{document}